\documentclass[dvipdfmx,12pt]{article} 
\usepackage{graphicx,amssymb,bbold,bm}
\usepackage{amsmath,amsfonts,epsfig}
\usepackage{braket}
\usepackage[usenames]{color} 
\usepackage{wrapfig}
\usepackage{amsmath,amsfonts,amssymb,bm}
\usepackage{amsfonts,bm}
\usepackage{tikz}
\usepackage{mathtools}

\newcommand{\p}{\partial}
\newcommand{\pslash}{p\kern-1ex /}
\newcommand{\qslash}{q\kern-1ex /}
\newcommand{\lslash}{l\kern-1ex /}
\newcommand{\sslash}{s\kern-1ex /}
\newcommand{\kaslash}{k_a\kern-2ex /}
\newcommand{\kbslash}{k_b\kern-2ex /}
\newcommand{\Dslash}{{\cal D}\kern-1.5ex /}

\newcommand{\bc}{\overline{c}}

\newcommand{\beqa}{\begin{eqnarray}}
\newcommand{\eeqa}{\end{eqnarray}}

\newcommand{\bpm}{\begin{pmatrix}}
\newcommand{\epm}{\end{pmatrix}}
\newcommand{\bbm}{\begin{bmatrix}}
\newcommand{\ebm}{\end{bmatrix}}

\def\p{\partial}

\makeatletter

\@addtoreset{equation}{section}
\makeatother
\usepackage{subfig}
\begin{document}


\voffset -0.7 true cm
\hoffset 1.5 true cm
\topmargin 0.0in
\evensidemargin 0.0in
\oddsidemargin 0.0in
\textheight 8.6in
\textwidth 5.4in
\parskip 9 pt
 
\def\Tr{\hbox{Tr}}
\newcommand{\be}{\begin{equation}}
\newcommand{\ee}{\end{equation}}
\newcommand{\bea}{\begin{eqnarray}}
\newcommand{\eea}{\end{eqnarray}}
\newcommand{\beas}{\begin{eqnarray*}}
\newcommand{\eeas}{\end{eqnarray*}}
\newcommand{\nn}{\nonumber}
\font\cmsss=cmss8
\def\C{{\hbox{\cmsss C}}}
\font\cmss=cmss10
\def\bigC{{\hbox{\cmss C}}}
\def\scriptlap{{\kern1pt\vbox{\hrule height 0.8pt\hbox{\vrule width 0.8pt
  \hskip2pt\vbox{\vskip 4pt}\hskip 2pt\vrule width 0.4pt}\hrule height 0.4pt}
  \kern1pt}}
\def\ba{{\bar{a}}}
\def\bb{{\bar{b}}}
\def\bc{{\bar{c}}}
\def\bphi{{\Phi}}
\def\Bigggl{\mathopen\Biggg}
\def\Bigggr{\mathclose\Biggg}
\def\Biggg#1{{\hbox{$\left#1\vbox to 25pt{}\right.\n@space$}}}
\def\n@space{\nulldelimiterspace=0pt \m@th}
\def\m@th{\mathsurround = 0pt}

\begin{titlepage}
\begin{flushright}
{\small OU-HET-1191}
 \\
\end{flushright}

\begin{center}

\vspace{5mm}
{\Large \bf {Flows of Rotating 
Extremal Attractor}} \\[3pt]   
{\Large \bf {Black Holes }}  \\[3pt] 

\vspace{6mm}

\renewcommand\thefootnote{\mbox{$\fnsymbol{footnote}$}}
Norihiro Iizuka${}^{1}$, 
Akihiro Ishibashi${}^{2}$ and 
Kengo Maeda${}^{3}$

\vspace{3mm}

${}^{1}${\small \sl Department of Physics, Osaka University} \\ 
{\small \sl Toyonaka, Osaka 560-0043, JAPAN}

${}^{2}${\small \sl Department of Physics and Research Institute for Science and Technology,} \\   
{\small \sl Kindai University, Higashi-Osaka 577-8502, JAPAN} 

${}^{3}${\small \sl Faculty of Engineering, Shibaura Institute of Technology,} \\   
{\small \sl Saitama 330-8570, JAPAN} 

\vspace{4mm}

{\small \tt 
{iizuka at phys.sci.osaka-u.ac.jp}, {akihiro at phys.kindai.ac.jp},   \\
{maeda302 at sic.shibaura-it.ac.jp}
}

\end{center}


\noindent

\abstract{We investigate the attractor mechanism of five-dimensional extremal rotating black holes in Einstein gravity minimally coupled with a multiplet complex scalar. 
By imposing regularity on the horizon, we show that the only possible attractor value of the scalar field is zero in our setup and that the local geometry is determined by the Myers-Perry black hole solution. We numerically obtain the extremal AdS black hole solutions interpolating the near horizon geometry to the asymptotic AdS spacetime under the existence of a bare potential of the scalar field. The black hole energy and the angular momenta are discretized for the usual Dirichlet boundary condition. Under the general boundary condition, we also find hairy extremal AdS black holes in which the energy is smaller than that of the extremal Myers-Perry AdS black hole solution for the same angular momentum.}

\end{titlepage}

\setcounter{footnote}{0}
\renewcommand\thefootnote{\mbox{\arabic{footnote}}}

\tableofcontents
\newpage

\section{Introduction} 
An attractor mechanism appearing in an extremal charged black hole is a beautiful phenomenon. 
A scalar field, that has no bare potential and therefore can take vacuum expectation value (VEV) freely at spatial infinity, can only take a specific value at the black hole horizon. Furthermore, the specific value at the horizon is determined by the charge of a black hole only. Thinking of the VEV of a scalar field as a function of radial coordinate,  the scalar VEV can take any value at infinity, but it is attracted to the specific value at the horizon.  
This attractor mechanism was found originally in ${\cal{N}}=2$ supersymmetric black holes \cite{Ferrara:1995ih}, and developed furthermore in \cite{Cvetic:1995bj, Strominger:1996kf, Ferrara:1996dd, Ferrara:1996um, Cvetic:1996zq, Ferrara:1997tw, Gibbons:1996af, Denef:2000nb, Denef:2001xn}
. Later  
it was shown in \cite{Goldstein:2005hq} to hold more generally at zero temperature without supersymmetry.

The fact that a scalar field can only take a specific value at the horizon for zero temperature black hole is deeply tied to the fact that near horizon geometry determines the entropy of the black hole. The entropy of the extremal charged black hole is determined by the charge only. Since the charge is quantized, its entropy should not allow any free parameter to tune. 
On the other hand, if there are moduli at the horizon, the existence of the free parameter at the horizon is in contradiction with the fact 
that the ground state entropy of the system is fixed by the charge only.

By thinking in this way, it is easy to imagine that the attractor phenomena works not only for an extremal charged black hole but also for a rotating extreme black hole. In fact, the attractor mechanism for rotating black holes has also been investigated \cite{Astefanesei:2006dd} where the attractor value is determined by extremizing the entropy function \cite{Sen:2005wa} and thus it is determined locally near the horizon. In many situations of rotating attractors, 
the extremality condition for black holes is achieved with nonzero charges. 
In this paper, we investigate the attractor mechanism of a rotating black hole without any charges. 

More concretely, we consider Einstein gravity minimally coupled to multi complex scalars, which is the model studied recently in \cite{Ishii:2021xmn}. 
In this case, the analytic nature of the horizon places strong restrictions on the scalar, forcing it to be zero under certain conditions which we clarify. Under this, in this paper, we investigate the rotating attractor mechanism and its deformation. 
This paper can be regarded as the  parallel work done here \cite{Iizuka:2022igv}, replacing the charge of a charged 
black hole with angular momentum. 

In general, it is much more difficult to analyze stationary rotating black holes than static ones   
since the geometry is no longer spherically symmetric, and the field equations become highly involved in partial differential equations. 
For the five-(more generally odd-)dimensions, however, there are exceptional cases in the sense that the Einstein equations coupled to a class of 
multi complex scalar fields reduce to ordinary differential equations when we apply the cohomogeneity-$1$ metric ansatz. 
The rotating boson star~\cite{Hartmann:2010pm} and rotating AdS black hole solutions~\cite{Dias:2011at} were constructed for complex doublet scalar models, on this ansatz.      
The complex scalar model was generalized to a multiplet complex scalar for both the rotating AdS black hole and 
rotating boson stars~\cite{Ishii:2021xmn}. In this paper, we investigate the attractor mechanism of extremal rotating 
black holes under the cohomogeneity-$1$ ansatz in the Einstein gravity minimally coupled with a multiplet 
complex scalar. We find that the only allowed attractor value is zero under certain conditions, due to the existence of 
angular momentum of the black hole. This means that the local geometry of the regular extremal AdS black 
hole is determined by the Myers-Perry black hole, as the scalar field rapidly decays to zero near the horizon. 

Introducing a relevant deformation to the complex scalar system through a bare potential,  we 
find various extremal rotating AdS black hole solutions interpolating near the horizon to the AdS boundary. 
In particular, we find that the mass and the angular momentum of the black hole solutions are 
discretized under the usual Dirichlet boundary condition at the AdS boundary. When the mass of the scalar field  
is negative, the scalar field admits two normalizable modes near the boundary. In this case, we find that the energy 
of the extremal rotating black hole with a complex multiplet scalar hair is less than that of the 
extremal Myers-Perry AdS black hole~\cite{Hawking:1998kw} for the same angular momentum. 
 
The organization of the paper is as follows. In section \ref{sec:2}, we briefly review the multiplet complex 
scalar model~\cite{Ishii:2021xmn} and investigate the attractor conditions for the extremal rotating black holes, where 
the near horizon geometry is analyzed. In section 3, we analyze asymptotically flat spacetime and in section \ref{sec:4}, we numerically find the extremal 
hairy black hole solutions interpolating the near horizon to the AdS boundary under the various boundary 
conditions. Section~\ref{sec:5} is devoted to summary and discussions.   

\section{Attractor mechanism for the near horizon geometry in extremal rotating black holes}
\label{sec:2} 
In this section, we investigate the near horizon geometry of five-dimensional extremal rotating black holes. 
Our model contains multiple complex scalar fields. 
We see that the attractor mechanism to work and the value at the horizon 
is strictly restricted to be zero. 
This shows that the local geometry of the extremal black holes is determined by the Myers-Perry black hole.   Note that the argument in this section \ref{sec:2} is local and independent of the asymptotic geometry.

\subsection{The setup}
We consider stationary extremal black hole solutions in the five-dimensional Einstein gravity minimally coupled 
to multi complex scalar fields $\vec{\Psi}$ with a potential $V(\vec{\Psi})$. The action is given by 
\begin{align}
\label{action}
S=\frac{1}{2}\int d^5x\sqrt{-g}\left[R-2|\nabla\vec{\Psi}|^2-2V(\vec{\Psi})\right],     
\end{align} 
where here and in the following we set $8\pi G=1$. 
We take the following metric ansatz:    
\begin{align}
\label{metric}
 ds_5^2 &=-f(u)g(u)dt^2+\frac{r_+^2}{4u^3f(u)}du^2+\frac{r_+^2}{4u}(d\theta^2+\sin^2\theta d\varphi^2) \nonumber \\
&\qquad +\frac{r_+^2}{u}h(u)\left(d\psi+\frac{\cos\theta}{2}d\varphi-\Omega(u)dt\right)^2, 
\end{align}
which is a cohomogeneity-1 spacetime. We choose the gauge that  
the horizon is located at $u=1$~($f(1)=0$). 
Each surface on $t= \mbox{const.}$ and $u= \mbox{const.}$ is homogenously squashed $S^3$, which is composed of $S^1$ fibers 
over $S^2$~(the angular coordinate ranges are $0\le \psi\le 2\pi$, $0\le \theta\le \pi$, and 
$0\le \varphi\le 2\pi$). Here, $r_+$ is the horizon raidus on $S^2$. When $h=1$, it reduces to a sphere $S^3$ with 
$SO(4)$ symmetry. This spacetime admits asymptotic timelike Killing vector $(\p_t)^\mu$ and spacelike vectors 
$\xi_i^\mu$ on $S^3$ given by Bianchi type IX,   
\begin{align}
& \xi_1=\cos\varphi\,\p_\theta+\frac{\sin\varphi}{2\sin\theta}\p_\psi-\cot\theta \sin\varphi\, \p_\varphi, \nonumber \\
& \xi_2=-\sin\varphi\,\p_\theta+\frac{\cos\varphi}{2\sin\theta}\p_\psi-\cot\theta \cos\varphi\, \p_\varphi, \nonumber \\
& \xi_3=\partial_\varphi. 
\end{align}
The null Killing vector on the horizon is given by the linear combination of $\p_t$ and $\p_\psi$ as  
\begin{align}
\label{Null_Killing}
l=\p_t+\Omega|_{u=1}\p_\psi, 
\end{align}
where $\Omega|_{u=1}$ is the angular velocity of the black hole.

For the case 
\begin{align}
V(\vec{\Psi}) =   \Lambda = - \frac{6}{\ell^2} 
\end{align}
{\it i.e.,} with negative cosmological constant, and in the absence of the scalar field, 
the action~(\ref{action}) admits  
the extremal Myers-Perry AdS black hole metric~\cite{Hawking:1998kw}
\begin{align}
\label{Myers-Perry}
& f(u)=\frac{r_+^2}{\ell^2u}(1-u)^2\left(1+2u+\frac{\ell^2}{r_+^2}u\right), \nonumber \\
& h(u)=1+\left(1+\frac{2r_+^2}{\ell^2}\right)u^2,\qquad g(u)=\frac{1}{h(u)}, \nonumber \\
& \Omega(u)=\frac{2r_+^2}{\ell^3}\left(1+\frac{\ell^2}{r_+^2}\right)\sqrt{1+\frac{\ell^2}{2r_+^2}}\frac{u^2}{h(u)} 
\end{align}
as a solution. 
Here $u=1$ is a horizon and $u=0$ is AdS boundary. 

From our ansatz for the metric eq.~\eqref{metric} and its Killing symmetry, all the equations reduce to a set of ordinary differential 
equations~(ODEs) if one considers $2j+1$ components of the scalar field in the form~\cite{Ishii:2021xmn} 
\begin{align}
\label{vector_form_scalar}
\vec{\Psi} &=\{\Psi_{j},\,\Psi_{j-1},\,\cdots,\,\Psi_{-j}  \} \nonumber \\
&=\phi(u)\,e^{-i\omega t}\left(\Theta^j_{j,k},\,\Theta^j_{j-1,k},\, \Theta^j_{j-2,k},\,\cdots,\,\Theta^j_{-j,k} \right) \,.
\end{align}
Here $\phi(u)$ is a real-valued function and $\Theta^j_{m,k}~(m=-j, -j+1, \cdots, j-1, j)$ are Wigner D-matrices, 
which are eigenfunctions of the Laplace equations on the homogeneously squashed $S^3$, and thus satisfy the following equations  
\begin{align}
\label{eigenfunction}
& \left(\sum_{i=1}^3L_i^2\right)\Theta^j_{m,k}=j(j+1)\Theta^j_{m,k}, \quad j=0, \,\frac{1}{2}, \,1,\,\frac{3}{2}, \cdots, 
\nonumber \\
& L_3\Theta^j_{m,k}=m\Theta^j_{m,k}, \quad m=-j,\,-j+1, \cdots,\,j. \nonumber \\
& \frac{i}{2}\p_\psi\Theta^j_{m,k}=k\Theta^j_{m,k}, \quad k=-j,\,-j+1, \cdots,\,j, 
\end{align}
where $L_i~(i=1,2,3)=i\xi_i$ is the angular momentum operator satisfying the commutation relation
\begin{align}
\label{comm_relation}
[L_i,\,L_j ]=i\epsilon_{ijk}L_k. 
\end{align}

Let us define the scalar potential $V(\vec{\Psi})$ as
\begin{align}
\label{def_potential}
& V(\vec{\Psi})=V(|\vec{\Psi}|)=-\frac{6}{\ell^2}+\frac{1}{2}M^2|\phi|^2+\cdots, \nonumber \\
& |\vec{\Psi}|\equiv \sqrt{\sum_{m=-j}^j\Psi_m \Psi^\ast_m}
=|\phi|\sqrt{\sum_{m=-j}^j\Theta^j_{m,k}\,{\Theta^\ast}^j_{m,k}  }
=|\phi|. 
\end{align}
Here $\cdots$ corresponds to terms of higher power in $|\phi|^2$ and 
we used the formula
\begin{align}
\label{summation_relation}
\sum_{m=-j}^j{\Theta^\ast}^j_{m,k} \Theta^j_{m,k}=1
\end{align} 
in the last equality. 
So, the scalar potential $V(\vec{\Psi})$ becomes the function of only $u$ via $|\phi(u)|$ in eq.~(\ref{vector_form_scalar}). 
Combining the properties of the Wigner D-matrix~(\ref{eigenfunction}), one can check that the stress-energy momentum tensor 
\begin{align}
\label{stress-energy}
& T_{\mu\nu}=\nabla_\mu \vec{\Psi}\cdot \nabla_\nu \vec{\Psi^\ast}
+\nabla_\mu \vec{\Psi^\ast}\cdot \nabla_\nu \vec{\Psi}
-g_{\mu\nu}(|\nabla\vec{\Psi}|^2+V(\vec{\Psi}))
\end{align}
is consistent with the metric ansatz eq.~(\ref{metric}) \cite{Ishii:2021xmn}. Although the complex scalar fields, eq.~(\ref{vector_form_scalar}), 
are functions of all the coordinates, $(t,\,u,\,\psi,\,\theta,\,\varphi)$, the backreaction of the complex $(2j + 1)$ scalars to the metric is determined only through $\phi$. 

From the Einstein equations $R_{\mu\nu}-(g_{\mu\nu}/2)R=T_{\mu\nu}$, we derive two coupled second order 
ODEs for $h(u)$ and $\Omega(u)$, and as  
\begin{align}
\label{eq_h}
& h''+\frac{3u\{f+h+(\epsilon_k^2+\epsilon_{k+1}^2)\phi^2\}+r_+^2V(|\phi|)-6u}{3u^2f}h'-\frac{h'^2}{h} \nonumber \\
& \qquad +\frac{2gh(1-h)+g\{4k^2-h(\epsilon_k^2+\epsilon_{k+1}^2)\}\phi^2+r_+^2uh^2\Omega'^2}{u^2fg}=0 \,, \\     
\label{eq_Omega}
& \Omega''-\left(\frac{1}{u}+\frac{g'}{2g}-\frac{3h'}{2h} \right)\Omega'-\frac{2k(\omega+2k\Omega)}{u^2fh}\phi^2=0 \,,
\end{align}
where the coefficient $\epsilon_{k}$ is defined by
\begin{align}
\label{def_epsilon_k}
\epsilon_k:=\sqrt{(j+k)(j-k+1)} 
\end{align}
and a dash denotes the derivative with respect to $u$. 
We also have two coupled first order ODEs for $f(u)$ and $g(u)$, which are given in Appendix~\ref{Appendix:A}. 

Due to the properties eq.~(\ref{eigenfunction}), 
the scalar field equation for $\Psi_m$ reduces to a single scalar equation, 
\begin{align}
\label{eq_phi}
& f\phi''+\left(f'+\frac{g'f}{2g}+\frac{h'f}{2h}  \right)\phi'+\left(r_+^2\frac{(\omega+2k\Omega)^2}{4u^3fg}
-\frac{k^2(1-h)}{u^2h}-\frac{j(j+1)}{u^2}  \right)\phi \nonumber \\
&\qquad -\frac{r_+^2\phi}{8u^3|\phi|}\frac{\partial V}{\partial |\phi|}=0.  
\end{align} 

\subsection{Attractor conditions}

Let us investigate the attractor value of the scalar field on the horizon. 
Hereafter, we assume that the metric is $C^2$ outside the horizon and the scalar field $\phi$ is $C^1$. Therefore the 
functions near the horizon can be expanded as 
\begin{align}
\label{series}
& \Omega= \Omega_0+\Omega_1(1-u)+\Omega_2(1-u)^2+\cdots, \nonumber \\
& h= h_0+h_1(1-u)+h_2(1-u)^2+\cdots, \nonumber \\
& f=f_0(1-u)^2+\cdots, \nonumber \\
& g=g_0+g_1(1-u)+\cdots, \nonumber \\
& \phi=\phi_0+\phi_1(1-u)^\lambda, \quad \lambda\ge 1,    
\end{align}
where the coefficients $h_0$, $f_0$, $g_0$ are positive, and the frequency $\omega$ must satisfy the condition 
\begin{align}
\label{corotating_cond}
\omega=-2k\Omega_0
\end{align}
so that the complex scalar field is stationary in a co-rotating frame near the extremal rotating black hole horizon. 
As shown in Ref.~\cite{Iizuka:2022igv}, in the presence of the bare potential $V(\phi)$, one can take any attractor values $\phi_0$ for the case of static charged extremal black hole by tuning the bare potential $V(\phi)$\footnote{Without bare potential, the attractor value $\phi_0$ is determined by the charges of the black hole only. See \S 3 in \cite{Iizuka:2022igv}}.  
In our rotating case, however, one cannot take arbitrary value $\phi_0$, and indeed   
$\phi_0=0$ is the only attractor value, as shown below.    

Suppose that $\phi_0\neq 0$. Then, substituting eqs.~(\ref{series}) into eq.~(\ref{eq_Omega}), one finds
\begin{align}
\label{cond_Omega_phi_0}
\Omega_1=0. 
\end{align} 
Let us consider a $t,\,u=const.$ spacelike surface $S_0$ on the future event horizon and a past-directed null geodesic congruence 
orthogonal to $S_0$ and transverse to the horizon. The tangent vector $k^\mu$ of affinely parametrized such null geodesics and the expansion $\theta$ thereof are given by 
\begin{align}
\label{outgoing_null}
& k^\mu=-E\left(\frac{1}{fg}\partial_t+\frac{\Omega}{fg} \partial_\psi+\frac{2u^\frac{3}{2}}{r_+\sqrt{g}}
\partial_u\right)^\mu, \nonumber \\
& \theta:=\frac{{\cal L}_k {\cal A}}{{\cal A}}=E\frac{3h-uh'}{r_+h}\sqrt{\frac{u}{g}},   
\end{align}  
where $E$ is a positive constant and ${\cal A}$ is the area element of the null geodesic congruence. 
In the usual black hole case, the expansion $\theta$ on $S_0$ needs to be positive in the outgoing direction. 
Thus, this leads to  
\begin{align}
\label{assum_expansion}
\theta(u=1)>0 \,\,\, \Longleftrightarrow \,\,\, h_1+3h_0>0 \,.
\end{align}


Under the condition~(\ref{corotating_cond}) and the expansion~(\ref{series}), eqs.~(\ref{eq_f}) and (\ref{eq_g}) 
are rewritten near the horizon as 
\begin{align}
\label{eq_fg_series}
& 3u^2(uh'-3h)f'=F_0+O(1-u), \nonumber \\
& 3u^2(uh'-3h)g'=\frac{g_0G_0}{f_0(1-u)^2}+O((1-u)^{-1}), 
\end{align} 
where the coefficients $F_0$ and $G_0$ are functions of $h_0$, $h_1$, $\phi_0^2$, and 
$V(|\phi_0|)$~(in details, see Appendix~\ref{Appendix:B}).  Since the metric 
functions are $C^2$ and $f'=0$ on the horizon, each coefficient $F_0$ and $G_0$ must be zero. 
This yields 
\begin{align}
\label{sol_phi0}
\phi_0^2=\frac{h_0(4-h_0-r_+^2V(|\phi_0|))}{2\{2k^2+h_0(\epsilon_k^2+\epsilon_{k+1}^2)\}}, \qquad h_1+3h_0 = 0 .  
\end{align} 
This, however, contradicts eq.~(\ref{assum_expansion}). Thus, $\phi_0$ must be zero in the extremal rotating black holes. 

\subsection{The local no-hair near the horizon}
\label{subsec:2.2}
As shown in the previous subsection, the attractor value of the scalar field $\phi_0$ must be zero for the 
extremal black holes. So we can expect that the back reaction of the scalar field on the metric 
is negligible near the horizon. To illustrate this, let us first consider the solution of the scalar field eq.~(\ref{eq_phi}) 
near the horizon on the background of the extremal Myers-Perry black hole metric~(\ref{Myers-Perry}).  
Expanding the scalar potential $V(|\phi|)$ as
\begin{align}
\label{potential_form}
V(|\phi|)=-\frac{6}{\ell^2}+\frac{1}{2}M^2|\phi|^2+O(|\phi|^3), 
\end{align}
eq.~(\ref{eq_phi}) near the horizon reduces to the following equation, 
\begin{align}
\label{near_ho_eq_Scalar}
& \qquad  (1-u)^2\phi''-2(1-u)\phi'-\zeta\phi\simeq 0, \nonumber \\
& \zeta=\frac{1}{1+3\hat{r}_+^2}\left[j(j+1)+\frac{\hat{r}_+^2\hat{M}^2}{8}-\frac{3k^2(1+2\hat{r}_+^2)}{2(1+3\hat{r}_+^2)} \right], 
\end{align}
and the solution is approximately given by 
\begin{align}
\label{alpha_pm}
\phi\simeq \alpha_+(1-u)^{\lambda_+}+\alpha_-(1-u)^{\lambda_-}, \qquad 
\lambda_\pm=\frac{-1\pm\sqrt{1+4\zeta}}{2},  
\end{align}
where here and hereafter 
\begin{equation}
\label{def_r_hat}
\hat{r}_+:= \dfrac{r_+}{\ell } \,, \quad \hat{M}:= \dfrac{M}{\ell} \,. 
\end{equation}  
By imposing that $\phi$ is $C^1$, the second term with index $\lambda_-$ should be discarded and 
$\zeta$ must satisfy the inequality 
\begin{align}
\label{zeta_inequality}
\lambda_+\ge 1\,\, \Longrightarrow \,\, \zeta\ge 2. 
\end{align}
This is equivalent to the inequality 
\begin{align}
j(j+1)\ge \frac{4(1+3\hat{r}_+)^2+3k^2(1+2\hat{r}_+^2)}{2(1+3\hat{r}_+^2)}-\frac{\hat{r}_+^2\hat{M}^2}{8} , 
\end{align}
which gives a lower bound of the index $j$ for each black hole horizon radius $r_+= \hat{r}_+ \ell $, the eigen value $k$, and 
the mass $M = \hat{M} \ell $ of the complex scalar field.  

Let us first look at the the back reaction on the metric function $\Omega$ 
in eq.~(\ref{eq_Omega}). The formal solution of eq.~(\ref{eq_Omega}) is given by 
\begin{align}
\label{sol_Omega}
\Omega'=u\sqrt{\frac{g}{h^3}}
\left[C+2k\int^u_1 \sqrt{\frac{h}{g}}\,\frac{\omega+2k\Omega}{u^3f}\,\phi^2du\right], 
\end{align}
where $C$ is an integration constant. Since $\lambda\ge 1$ in eqs.~(\ref{series}), 
the scalar field appears in $\Omega'$ in the following subleading terms as  
\begin{align}
\label{scalar_Omega'}
\Omega'= u\sqrt{\frac{g}{h^3}}\left(C+ B (1-u)^{2\lambda}\right),  
\end{align} 
where $B$ is some coefficient, whose explicit expression is not relevant for the rest of our argument. 
Similarly, it is expected that the back reaction effect of the scalar field on the other metric functions 
are also negligible near the horizon. In other words, the near horizon geometry of the extremal 
black hole locally approaches AdS$_2$ geometry. Hereafter, we will show that the geometry near the horizon 
is given by the Myers-Perry extremal black hole metric~(\ref{Myers-Perry}) apart from local gauge freedom.  

Substituting eqs.~(\ref{series}) and (\ref{scalar_Omega'}) into eq.~(\ref{eq_h}), we obtain 
\begin{align}
\label{constr1}
\dfrac{\{uh-2(\hat{r}_+^2+u)\}hh'+2h^2(1-h) + {\ell^2C^2\hat{r}_+^2u^3} }{u^2f h} 
+h''-\frac{h'^2}{h} 
+\frac{h'}{u}=O((1-u)^{2\lambda-2}). 
\end{align}
Since the metric is $C^2$, we obtain the numerator of the first term in eq.~(\ref{constr1}) should be 
$O((1-u)^2)$, i.~e.~,   
\begin{align}
\label{constr1}
\{uh-2(\hat{r}_+^2+u)\}hh'+2h^2(1-h)+\ell^2C^2\hat{r}_+^2u^3=O((1-u)^2). 
\end{align}
Similarly, by requiring $f'=O(1-u)$, we obtain the other constraint from eq.~(\ref{eq_f}) as 
\begin{align}
\label{constr2}
\ell^2C^2\hat{r}_+^2u^4+h^2(6\hat{r}_+^2+8u-5uh)+2uhh'(uh-2\hat{r}_+^2-2u)=O(1-u). 
\end{align} 
Substituting the expansion~(\ref{series}) into eqs.~(\ref{constr1}) and (\ref{constr2}), 
we obtain the coefficients $h_0$, $h_1$, and $C^2$ under the condition~(\ref{assum_expansion})
as 
\begin{align}
\label{Coefficiengt_h}
& h_0=2\left(1+\hat{r}_+^2\right), \qquad h_1=-2(1+2\hat{r}_+^2), \nonumber \\
& C^2=\frac{8(1+2\hat{r}_+^2)(1+\hat{r}_+^2)^2}{\ell^2\hat{r}_+^2}. 
\end{align}
This is nothing but the coefficient of the metric expansion of the Myers-Perry black hole~(\ref{Myers-Perry}). By 
calculating the next subleading order in eq.~(\ref{eq_f}), we also obtain $f_0=1+3\hat{r}_+^2$, agreeing with the 
metric expansion of the Myers-Perry black hole~(\ref{Myers-Perry}). 
It is noteworthy that the metric~(\ref{metric}) 
and the field equations~(\ref{eq_h}), (\ref{eq_Omega}), (\ref{eq_f}), (\ref{eq_g}), and (\ref{eq_phi}) are invariant 
under the transformation:
\begin{align}
\label{transform}
\psi\to \psi+\xi\,t, \qquad \Omega_0\to \Omega_0+\xi, \qquad \omega\to \omega-2k\xi. 
\end{align} 
So, apart from this gauge freedom and the scaling of $t$, any regular solution 
under our ansatz for the metric and the complex scalar field reduces to the Myers-Perry black hole~(\ref{Myers-Perry}) locally near the horizon. 

\section{Asymptotically flat case}
Before considering the asymptotically AdS spacetime, it is worth noting that for generic $\omega$, the solution cannot be asymptotically flat. 
Suppose the asymptotically flat solution would be realized by taking the limit $\ell \to \infty$. 
Then, 
\begin{align}
f=g=h=1, \qquad \Omega=0 \,\,\, \mbox{for} \,\,\, u=0, 
\end{align}
and the asymptotic form of eq.~(\ref{eq_phi}) becomes 
\begin{align}
\label{phi}
\phi''\simeq -\frac{a^2}{4u^3}\phi \qquad a^2:=r_+^2\left(\omega^2-\frac{M^2}{2}\right)
\end{align}
in the asymptotic region, $u\to 0$. 
The asymptotic solution is given by the (modified) Bessel functions $J_\nu(x)~(I_\nu(x))$ and 
$N_\nu(x)~(K_\nu(x))$, depending on the sign of $a^2$ as   
\begin{align}
\label{phi_flat_sol}
& \phi\simeq \sqrt{u}\left[c_1J_1\left(\frac{a}{\sqrt{u}}  \right)
+c_2N_1\left(\frac{a}{\sqrt{u}}  \right)     \right], \qquad a^2>0, \nonumber \\
& \phi\simeq \sqrt{u}\left[ {\tilde c}_1I_1\left(\frac{\tilde{a}}{\sqrt{u}}  \right)
+ {\tilde c}_2K_1\left(\frac{\tilde{a}}{\sqrt{u}}  \right)     \right], \qquad 
\tilde{a}^2:=-a^2>0. 
\end{align}
Let us analyze the following three cases separately. 
\begin{enumerate}
\item $\omega^2<M^2/2$ case: the solution is given by the linear combination of the 
Bessel functions. By imposing the regularity condition on the horizon, generically the solution includes the asymptotically diverging mode, $I_1(\tilde{a}/\sqrt{u})\sim e^{\tilde{a}/\sqrt{u}}\to \infty$. 
Thus in the case of $\omega^2<M^2/2$, for the generic value of $\omega$, there is no asymptotically flat solution for the $\omega$ satisfying the co-rotating condition eq.~(\ref{corotating_cond})\footnote{However, for some quantized number $\omega$~(or $\Omega_0$ in Eq.~(\ref{corotating_cond})), 
the regularity condition both on the horizon and at the asymptotic infinity can be satisfied. Thus  
there may possibly be an exceptional case in which one can adjust the co-rotating condition,  (\ref{corotating_cond}), for that quantized number $\omega$, so that the asymptotically flat solution can be the case.  We will not investigate such a spacial case here anymore.}.  
\item $\omega^2=M^2/2$ case: the solution is given from eq.~(\ref{phi}) as,  
\begin{align}
\phi \simeq {c}'_1  + \frac{{c}'_2}{r^2} \,, \quad \mbox{near $u=0$}
\end{align}  
where we used $u=r_+^2/r^2$. 
The regularity condition imposed at the horizon requires specific linear combinations between $c_1'$ and $c_2'$, and generically the constant $c_1' \neq 0$. However, when $\omega \neq 0$ due to the co-rotating condition eq.~(\ref{corotating_cond}), this implies that the gravitational energy diverges due to its oscillation at infinity with nonzero magnitude, and thus, an asymptotically flat solution is not allowed. 
\item $\omega^2 > M^2/2$ case:  the asymptotic behavior of $\phi$ is given by 
\begin{align}
\label{asymptotic_form_flat}
\phi\sim u^{3/4}\cos\left(\frac{a}{\sqrt{u}}+\delta  \right) \sim {O}\left({r^{-3/2}}\right) \,.
\end{align} 
However, this scalar field does not decay fast enough asymptotically as $\phi \sim r^{-2}$ which follows the Gauss law. In other words, 
by substituting Eq.~(\ref{asymptotic_form_flat}) into eq.~(\ref{eq_f}) 
\begin{align}
f\simeq 1+O(\sqrt{u})\simeq 1+ O\left(\frac{1}{r}\right),  
\end{align}
Thus, we conclude that for the generic value of $\omega$, 
the gravitational energy is divergent for such slow decay of the scalar field, and thus one cannot have an asymptotically flat solution as $f\simeq 1+O(1/r^2)$.  
\end{enumerate}
Thus unlike the static extremal attractor black holes, in the rotating extremal black hole case,  due to the co-rotating condition eq.~(\ref{corotating_cond}), the construction of asymptotically flat black holes in our setup is highly restricted and impossible generically. Therefore in the next section, we consider the case with asymptotically AdS case. 

\section{The attractor black hole solutions in AdS$_4$} 
\label{sec:4} 
In this section, we numerically construct 
extremal AdS black hole solutions with complex scalars 
interpolating between the near horizon solution~(\ref{near_ho_eq_Scalar}) and asymptotically AdS solution. 
In our setup, we consider the case that the mass-squared, $M^2$, of the scalar field is negative, 
and that the metric asymptotically approaches AdS spacetime~\footnote{If one introduces a radial 
coordinate $r$ by $u=r_+^2/r^2$, the metric becomes usual asymptotic form, 
$ds^2\simeq -r^2/\ell^2dt^2+\ell^2/r^2dr^2+r^2d\Omega_3^2$.}, 
\begin{align}
\label{asymptotic_AdS}
f\simeq \frac{r_+^2}{\ell^2 u}, \qquad h\simeq g\simeq 1, \qquad  \Omega\simeq 0. 
\end{align}
Then, the asymptotic solution of Eq.~(\ref{eq_phi}) is 
\begin{align}
\label{asym_phi}
\phi=k_+u^{\rho_+}+k_-u^{\rho_-}, \qquad \rho_\pm=1\pm \sqrt{1+\frac{\hat{M}^2}{8}} .  
\end{align}
Since both of the modes are normalizable, i.~e.,~$\rho_\pm>0$, $\phi$ approaches 
zero toward the AdS boundary, $u=0$.

As shown in the previous section, the attractor value of the scalar field is zero and the 
near horizon geometry reduces to that of the Myers-Perry black hole via  
eq.~(\ref{Coefficiengt_h}), and $f_0=1+3\hat{r}_+^2$, except the gauge freedom~(\ref{transform}). 
This gauge freedom~(\ref{transform}) is used to satisfy one of the asymptotic boundary 
conditions~(\ref{asymptotic_AdS}), $\Omega(0)=0$. Another trivial gauge freedom is the choice of $g_0$ in 
eqs.~(\ref{series}), corresponding to the scaling of $t$ as $t\to s_0 t$ with some constant $s_0 \neq 0$. 
This freedom is used to satisfy another asymptotic boundary condition, $g(0)=1$ in the boundary 
conditions~(\ref{asymptotic_AdS}). So, once the coefficients $h_0$, $h_1$, $\Omega_1$, and $f_0$ in 
the expansion~(\ref{series}) are determined, the ODEs~(\ref{eq_h}), (\ref{eq_Omega}), (\ref{eq_f}), (\ref{eq_g}) are 
uniquely determined except the gauge freedoms mentioned above. 
So, the remaining free parameters on the horizon are  
\begin{align}
\label{free_parameter}
\alpha_+, \qquad \hat{r}_+
\end{align}
in Eqs.~(\ref{alpha_pm}) and (\ref{def_r_hat}). 
For simplicity, we consider the following potential with two free parameters, 
$b_1$ and $b_2$ as 
\begin{align}
\label{potential_form1}
V(|\phi|)=-\frac{6}{\ell^2}-\frac{15}{2\ell^2}|\phi|^2+\frac{b_1}{\ell^2}|\phi|^3+\frac{b_2}{\ell^2}|\phi|^4, 
\end{align}   
where $\hat{M}^2=-15/2$. In this case, the asymptotic solution~(\ref{asym_phi}) reduces to 
\be 
\label{asy_scalar_form}
\phi\simeq k_-u^{3/4}+k_+u^{5/4} ,   
\ee
where each mode is normalizable.

First, we numerically find the regular extremal black hole solutions under the usual 
boundary condition 
\be
\label{Dirichlet_bc}
k_-=0. 
\ee
\begin{figure}[htbp]
 \begin{minipage}{0.5\hsize}
  \begin{center}
   \includegraphics[width=70mm]{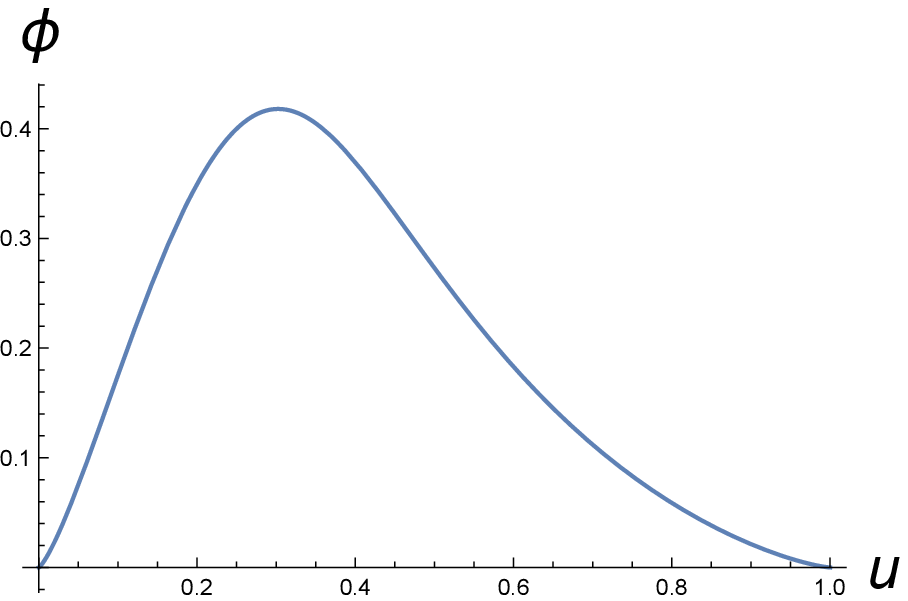}
  \end{center}
  \caption{\small{$\phi$ with $\hat{r}_+\simeq 2.91$, $j=9$, 
  $k=-1$ in the odd potential.}}
  \label{Odd_phi_j=9}
 \end{minipage}
\begin{minipage}{0.5\hsize}
  \begin{center}
   \includegraphics[width=70mm]{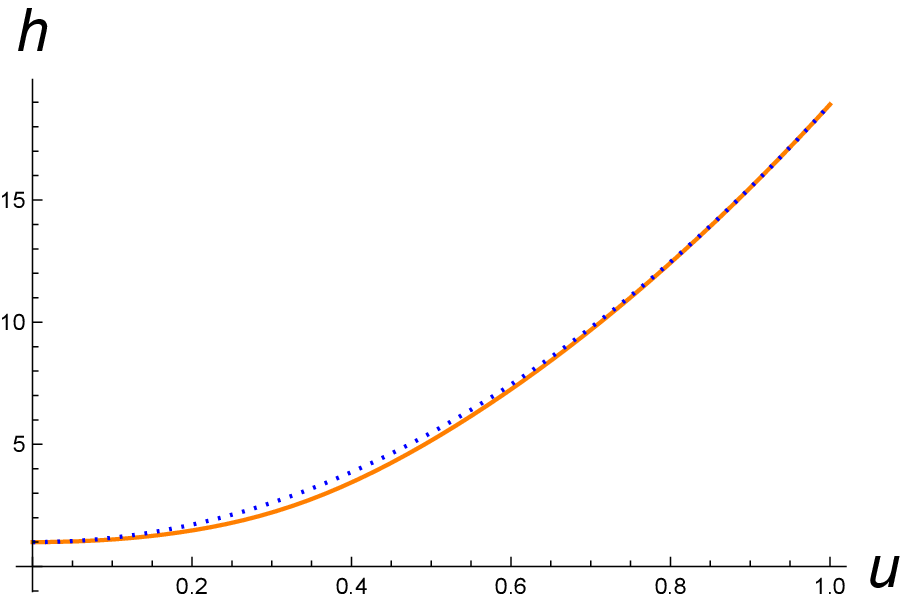}
  \end{center}
  \caption{\small{$h$ with $\hat{r}_+\simeq 2.91$, $j=9$, and $k=-1$ in the odd potential. 
  The dotted (blue) curve represents Myers-Perry extremal BH. }}
  \label{Odd_h_j=9}
 \end{minipage}
 \begin{minipage}{0.5\hsize}
  \begin{center}
   \includegraphics[width=70mm]{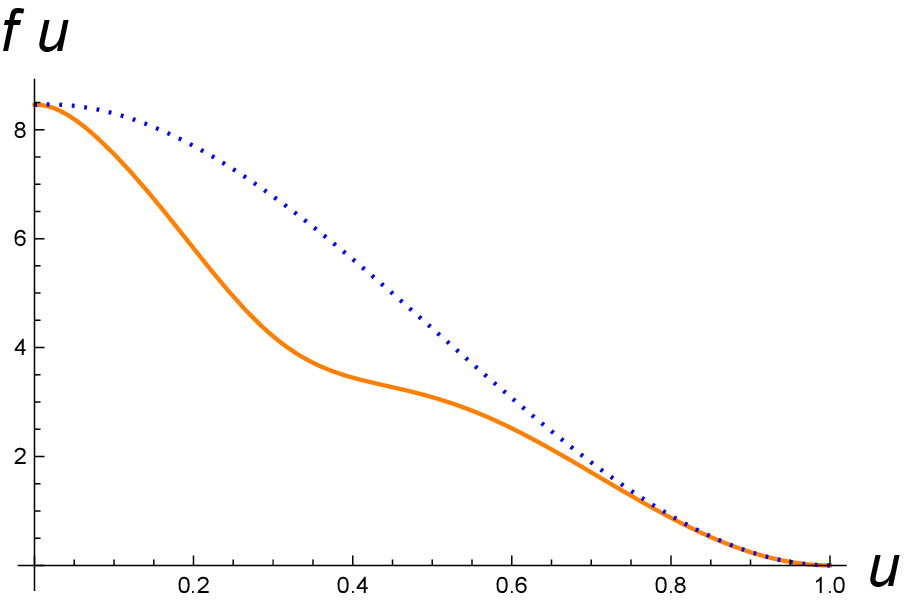}
  \end{center}
  \caption{\small{$fu$ with $\hat{r}_+\simeq 2.91$, $j=9$, and $k=-1$ in the odd potential. 
  The dotted (blue) curve represents Myers-Perry extremal BH. }}
  \label{Odd_f_j=9}
 \end{minipage}
 \begin{minipage}{0.5\hsize}
  \begin{center}
   \includegraphics[width=70mm]{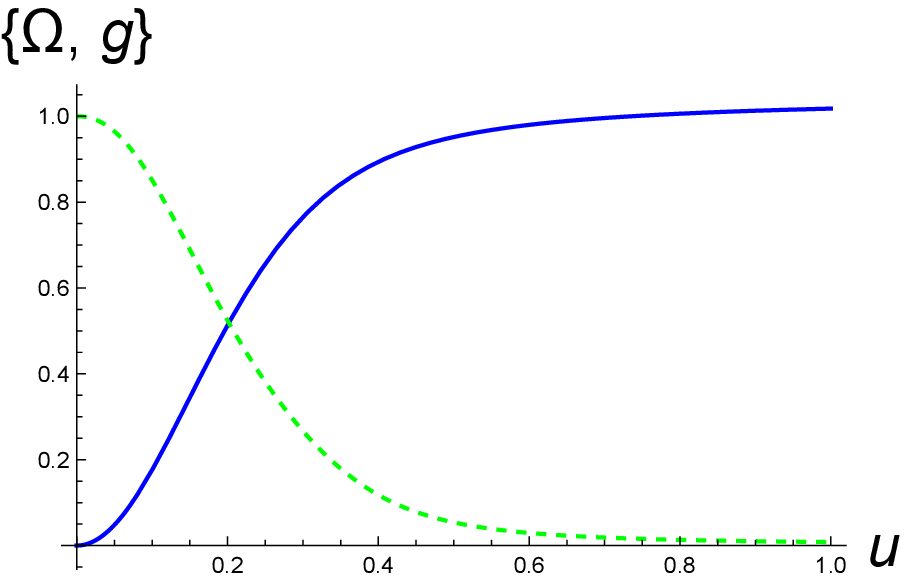}
  \end{center}
  \caption{\small{$(\Omega, g)$ with $\hat{r}_+\simeq 2.91$, $j=9$, and $k=-1$ in the 
  odd potential. The solid (blue) and dashed (green) curves are $\Omega$ and $g$, respectively. }}
  \label{Odd_Omega_j=9}
 \end{minipage}
\end{figure}

Among the free parameters $\Omega_0$, $g_0$, $\alpha_+$, and $\hat{r}_+$ on the 
horizon~(\ref{free_parameter}), the first two parameters are adjusted to satisfy 
the asymptotic boundary conditions, $g(0)=1$ and $\Omega(0)=0$ in (\ref{asymptotic_AdS}).   
Then, the remaining boundary conditions are Eq.~(\ref{Dirichlet_bc}) and 
\begin{align}
h(0)=1 
\end{align}
for the only two free parameters, $\alpha_+$ and 
$\hat{r}_+$~\footnote{$f\simeq r_+^2/u$ in  (\ref{asymptotic_AdS}) is automatically satisfied under the 
boundary condition~(\ref{asy_scalar_form}).}. This implies that the solutions are discretized for 
each ``quantum'' numbers, $j$ and $k$. In other words, the horizon radius and the entropy are 
characterized by such quantum numbers of the complex scalar field. Since it is in general difficult 
to find numerical solutions interpolating between the horizon and the AdS boundary in the full 
non-linear regime $|\phi|\gg 1$, we shall pay attention to the perturbed region in which 
$|\phi|\lesssim 1/2$ is satisfied and the deviation from the extremal Myers-Perry black hole 
solution is not so large.  

In the perturbation region, $|\alpha_+|\lesssim 1/2$, we find two discrete solutions for 
each $(j,\,k)=(7, -1), \,(9,-1)$ in the odd potential case, $b_1=-23$ and $b_2=0$. 
The numerical data for $(j,\,k)=(9,-1)$ are given in Figs.~\ref{Odd_phi_j=9}, \ref{Odd_h_j=9}, 
\ref{Odd_f_j=9}, \ref{Odd_Omega_j=9}. We find two discrete solutions with horizon radii, 
$\hat{r}_+=2.91,\, 3.07$ for $(j,\,k)=(9,-1)$. 
Similarly, we also find two discrete solutions with $\hat{r}_+=2.5,\, 2.7$ for $(j,\,k)=(7,-1)$. 

Secondly, we investigate regular solutions for generalized boundary conditions with $k_\pm\neq 0$.   
In this case, the total energy generically diverges according to the slower fall off 
term with coefficient $k_-$in Eq.~(\ref{asy_scalar_form}). To avoid such a divergence, we 
follow the approach adopted by Ref.~\cite{Hertog:2004ns}.   

To calculate the energy, we use the usual radial coordinate $r=r_+/\sqrt{u}$ instead of dimensionless 
coordinate $u$. By Eqs.~(\ref{eq_h}), (\ref{eq_Omega}), (\ref{eq_phi}), (\ref{eq_f}), and (\ref{eq_g}), one 
obtains the asymptotic solution near the AdS boundary as 
\begin{align}
\label{Near_bd_sol}
& f=\frac{r^2}{\ell^2}+1+\frac{k_-^2r_+^3}{\ell^2r} + \frac{C_1}{r^2}+\cdots, \nonumber \\
& g=1-\frac{k_-^2r_+^3}{r^3} + \cdots \nonumber \\
& h=1+ \frac{C_2}{r^4}+\cdots, \nonumber \\
& \Omega= \frac{C_3}{r^4}+\cdots,  
\end{align}
with $C_1, C_2, C_3$ being some constants. 
Due to the existence of the third term for $f$ in the asymptotic expansion~(\ref{Near_bd_sol}), usual gravitational 
energy diverges. However, as shown below, the total Hamiltonian of the gravitational energy and the scalar field 
energy is finite. 

According to the Hamiltonian approach, let us consider a timelike vector field $\xi=\partial_t$. 
Then, the variation of the gravitational energy is given by 
\begin{align}
\label{energy_G} 
& \delta Q_G[\xi]=\frac{1}{2}\int d\psi d\theta d\varphi\, 
\delta^r_i\,\overline{G}^{ijkl}(\xi_\perp \bar{D}_j\delta h_{kl}-\delta h_{kl}\bar{D}_j \xi^\perp), 
\nonumber \\
& G^{ijkl}=\frac{1}{2}\sqrt{g}(g^{ik}g^{jl}+g^{il}g^{jk}-2g^{ij}g^{kl}), \qquad h_{ij}=g_{ij}-\bar{g}_{ij}. 
\end{align} 
Here, $\bar{g}_{ij}~(i,\,j=r, \psi,\theta, \varphi)$ is the spatial metric of pure AdS and any geometric quantity $A$ with respect to 
$\bar{g}_{ij}$ is denoted by $\bar{A}$, and $\xi_\perp=\xi^\mu n_\mu$, with $n^\mu$ being the (past-directed) unit normal to 
the Cauchy surface $\Sigma$. 
The variation of the surface term by the scalar 
field is also given by   
\be 
\label{energy_scalar}
\delta Q_\phi=-2\int \xi^\perp \delta\phi D_i\phi \,dS^i=
-\frac{1}{2}\int \xi^\perp \delta\phi D_i\phi \,\delta^i_r\,r^3\sqrt{fh}\sin\theta \,d\psi d\theta d\varphi. 
\ee
In general, both eqs. \eqref{energy_G} and \eqref{energy_scalar} diverge, due to the slower fall-off of the scalar 
field with (\ref{asy_scalar_form}). 
Under the asymptotic expansion~(\ref{Near_bd_sol}), one can evaluate $\delta Q_G[\xi]$ 
and $\delta Q_\phi[\xi]$ as
\begin{align}
& \delta Q_G=\pi^2\left(-\frac{3\delta (\alpha^2)}{\ell^2}r
-3\delta C_1+\frac{4\delta C_2}{\ell^2}    \right), \nonumber \\
& \delta Q_\phi=\frac{\pi^2}{\ell^2}\left(3\delta (\alpha^2)r+6\delta (\alpha\beta)
+4\beta \delta\alpha \right), 
\end{align}
where $\xi_\perp=\sqrt{fg}$, and $\alpha$, $\beta$ are defined by 
\begin{align}
\alpha=k_-r_+^{3/2}, \qquad \beta=k_+r_+^{5/2}. 
\end{align}
Since the total variation of the gravitational and scalar charge is finite, 
we can define the total energy by choosing a function $\beta(\alpha)$ and introducing an effective 
potential $W(\alpha)$ following \cite{Witten:2001ua, Hertog:2004ns} as  
\begin{align}
\label{effective_potential}
W(\alpha)=\int^\alpha_0 \beta(\alpha)d\alpha. 
\end{align}
So, the total charge is finite and it is given by 
\begin{align}
\label{Q_total}
Q_{tot}=\pi^2\left(\frac{4C_2}{\ell^2}-3C_1+\frac{6\alpha \beta}{\ell^2}+\frac{4W(\alpha)}{\ell^2} \right). 
\end{align}
\begin{figure}[htbp]
  \begin{center}
   \includegraphics[width=110mm]{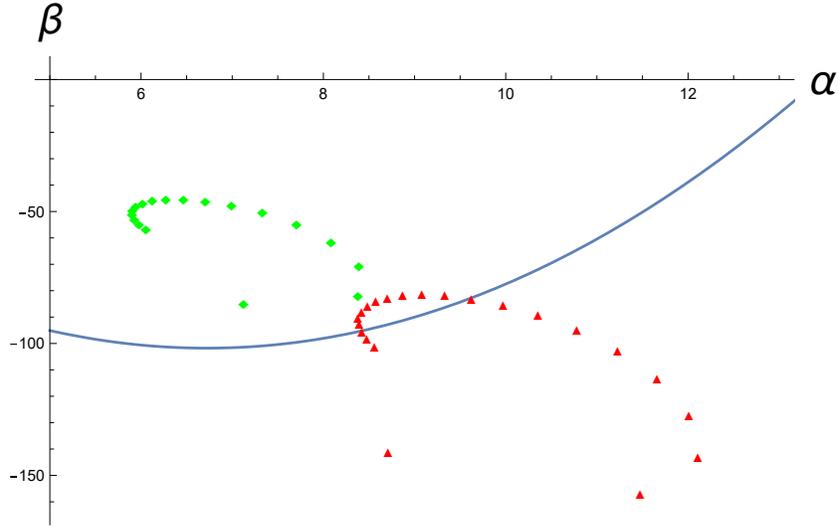}
  \end{center}
  \caption{\small{$\alpha$-$\beta$ plot for $(j,\,k)=(7,-1)$~(green diamonds), $(9, -1)$~(red triangles) for 
   various radii $2.48\le \hat{r}_+\le 2.82$~($j=7$), $3.0\le \hat{r}_+\le 3.42$~($j=9$)
    in the even potential with $b_1=0$, $b_2=-63$. The blue curve 
    $\beta\simeq 2.25\alpha(\alpha-13.4)$ intersects two solutions in $j=9$.}}
  \label{alpha_beta_plot_even}
\end{figure}
\begin{figure}[htbp]
  \begin{center}
   \includegraphics[width=110mm]{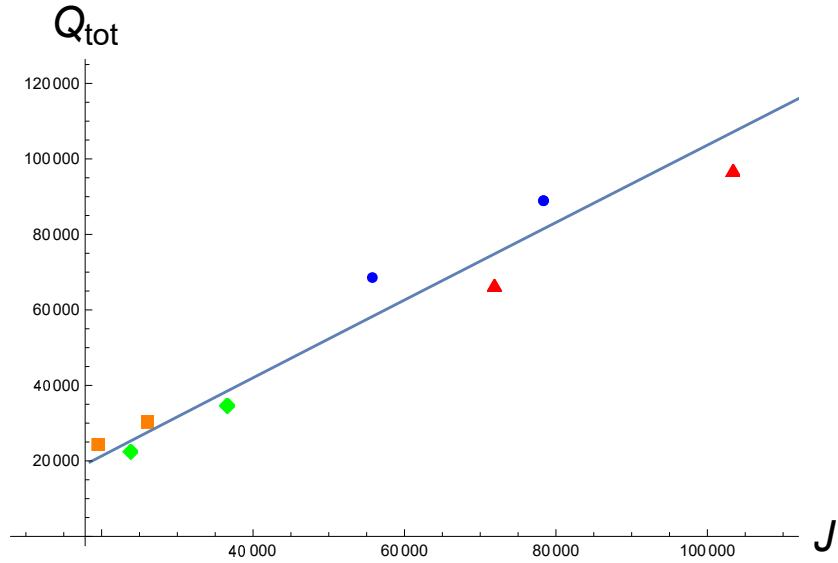}
  \end{center}
  \caption{\small{$(Q_{tot},\,J)$ plot.  The curve corresponds to extremal Myers-Perry black holes. 
  The (orange) boxes and (blue) circles represent $(j,\,k)=(7,\,-1)$ and $(9,\,-1)$ solutions satisfying 
  the boundary condition~(\ref{Dirichlet_bc}) for the odd potential $(b_1,\,b_2)=(-23,0)$. The (green) diamonds 
  and (red) triangles represent $(j,\,k)=(7,\,-1)$ and $(9,\,-1)$ solutions satisfying 
  the generalized boundary conditions~(\ref{dual_theory}) with 
  $(k,\,\alpha_0)\simeq (2.34,\,9.90)$ and  $(2.25,\,13.43)$.      }}
  \label{MJ_plot}
\end{figure}

The angular momentum can be read off from the coefficient $C_3$ in the expansion~(\ref{Near_bd_sol}). 
Since the falloff of $\Omega$ is the same as that of the Myers-Perry black hole~(\ref{Myers-Perry}), 
it is calculated by 
\begin{align}
\label{angular_momentum}
J=\frac{1}{2}\int_{\cal S} \epsilon_{\mu\nu \alpha\beta\gamma}\nabla^\beta\,\psi^\gamma=4\pi^2C_3, 
\end{align} 
where ${\cal S}$ is the three-dimensional sphere at the AdS boundary and $\psi^\mu=(\partial_\psi)^\mu$ is 
the axial Killing vector~\footnote{Eq.~(\ref{angular_momentum}) is normalized so that 
extremal Myers-Perry black hole satisfies the first law of the black hole for the gravitational 
energy~(\ref{Q_total}) and the angular momentum~(\ref{angular_momentum}). See Appendix~\ref{Appendix:C} in detail.}.

Before evaluating the energy and the angular momentum of the rotating black holes with a complex multiplet 
scalar hair, let us look at the 
solutions in $(\alpha,\,\beta)$ space in the even potential case, $b_1=0$. 
The figure~\ref{alpha_beta_plot_even} shows $(j,\,k)=(7,\,-1)$, $(9,-1)$ solutions for $\ell=1$ in $\alpha-\beta$ 
plane for various radii $\hat{r}_+$ in the even potential with $b_2=-63$ in the perturbative 
region $|\alpha_+|\lesssim 1/2$. Now, let us choose our boundary condition as 
\begin{align}
\label{dual_theory}
\beta=k\alpha(\alpha-\alpha_0), 
\end{align}   
where $k$ and $\alpha_0$ are some positive constants. When $\alpha$ is small enough, this corresponds to the 
usual Robin~(mixed) boundary condition, and for large $\alpha$ it reduces to $\beta\simeq k\alpha^2$, 
preserving all the asymptotic AdS symmetries~\cite{Hertog:2004dr}. 
The solid curve in Figure~\ref{alpha_beta_plot_even} corresponds to $k\simeq 2.25$ and $\alpha_0\simeq 13.4$. 
So, there are three solutions 
for the theory; two hairly extremal black hole solutions with small $\alpha$ and large $\alpha$, and $\alpha=0$ 
extremal Myers-Perry AdS solution~(\ref{Myers-Perry}).

We calculate the total energy~(\ref{Q_total}) and 
the angular momentum~(\ref{angular_momentum}) and compare them with those of 
the extremal Myers-Perry AdS black hole~(\ref{Myers-Perry}). 
The figure~\ref{MJ_plot} shows $(Q_{tot},\,J)$ plot for the several hairly 
extremal black hole solutions. The solid curve corresponds to the one-parameter family of extremal Myers-Perry 
AdS black hole solutions. The left two boxes above the curve correspond to the two solutions with $(j,\,k)=(7,\,-1)$ for 
the usual boundary condition~(\ref{Dirichlet_bc}), while the right two circles above the curve 
do to the ones with $(j,\,k)=(9,\,-1)$ for the same boundary condition~(\ref{Dirichlet_bc}). These imply 
that the energy of the hairly extremal black hole solution is larger than that of the extremal Myers-Perry black hole with the same angular momenta. The diamonds below 
the curve correspond to the two solutions with $(j,\,k)=(7,\,-1)$ for 
the generalized boundary condition~(\ref{dual_theory}) with $(k,\,\alpha_0)\simeq  (2.34,\,9.90)$, 
while the right two triangles below the curve 
do to the ones with $(j,\,k)=(9,\,-1)$ for the boundary condition $(k,\,\alpha_0)\simeq (2.25,\,13.43)$. 
The energies of these hairly extremal black hole solutions are less than those of the extremal 
Myers-Perry black holes with the same angular momenta.

\section{Summary and discussions}
\label{sec:5} 
We have investigated the attractor mechanism of the extremal rotating black holes in the Einstein gravity 
minimally coupled with a multiplet complex scalar field. In the non-rotating Einstein-Maxwell-dilaton 
system~\cite{Iizuka:2022igv}, one can take any attractor values by adjusting the bare potential. On the 
other hand, the only attractor value is zero in the extremal rotating case. This is because the non-zero value 
is incompatible with the regularity of the extremal horizon, as shown in section~\ref{sec:2}. So, the scalar 
field must decay to zero rapidly near the horizon, and the near horizon geometry coincides with the Myers-Perry 
AdS black hole~\cite{Hawking:1998kw}, as described in subsection~\ref{subsec:2.2}.  Note that the argument in section \ref{sec:2} is local and independent of the asymptotic region. 
This is similar to the local argument of the entropy function extremization  \cite{Sen:2005wa}. 
  
Then we discussed that this extremal rotating black hole does not admit asymptotic flat spacetime.   The crucial difference from the static extremal black hole case is that due to the co-rotating condition \eqref{corotating_cond}, the scalar field is oscillating and it does not admit finite energy configuration. 
 
We have also investigated the extremal rotating AdS black hole solution interpolating the near horizon 
geometry of the Myers-Perry black hole to the AdS boundary. Since the near horizon geometry is 
restricted to the Myers-Perry metric, the free parameters near the horizon are the magnitude 
$\alpha_+$ of the scalar field in Eq.~(\ref{alpha_pm}) and the horizon radius $\hat{r}_+$ only. If one 
imposes usual Dirichlet boundary condition for the scalar field in addition to the 
asymptotic boundary condition~(\ref{asymptotic_AdS}), two boundary conditions must be imposed at the 
AdS boundary. This implies that the extremal hairly black hole solution exists for a discrete set of the 
parameters $(\alpha_+,\, \hat{r}_+)$. In other words, the energy and the angular momentum are 
discretized, according to the quantum numbers of the scalar field, $(j,\,k)$ and the parameters 
$(b_1,\,b_2)$ of the bare potential~(\ref{potential_form1}). In this sense, all the properties of the 
extremal black hole are determined by the ``quantum'' state of the scalar field.  
These are quite different from the Myers-Perry AdS black hole in which the solution exists continuously 
by varying black hole radius $r_+$. 
 
For more general boundary condition~(\ref{asy_scalar_form}) with $k_\pm\neq 0$, we can evaluate 
the energy and the angular momentum of the extremal rotating black hole. As shown in section~\ref{sec:4}, 
there are hairly rotating extremal AdS black hole solutions whose energy is lower than that of the extremal Myers-Perry 
AdS black hole for the same angular momentum. This implies that the extremal Myers-Perry 
AdS black hole is unstable against the general boundary condition, as already observed in the static AdS 
black hole case~\cite{Hertog:2005hu}. As shown in~\cite{Dias:2011at, Ishii:2021xmn}, the Myers-Perry 
AdS black hole is unstable due to superradiance when the black hole horizon is small compared with 
the AdS scale $\ell$. In our case, the horizon radius is much larger than the AdS scale, i.~e.~, $\hat{r}_+\gg 1$, the instability 
of the extremal Myers-Perry AdS black hole would indicate that the vaccum state with zero expectation value 
of the scalar operator ${\cal O}$ dual to the complex scalar fields $\vec{\Psi}$ is unstable for the dual deformed 
boundary field theory. 
It would be interesting to pursue further the issue for a wide range of quantum numbers, $(j,\,k)$, the parameters 
of the potential, and the black hole radius.

\bigskip
\goodbreak
\centerline{\bf Acknowledgments}
\noindent

This work was supported in part by JSPS KAKENHI Grant No. 18K03619 (N.I.), 15K05092, 20K03938 (A.I.), 20K03975 (K.M.) and also supported by MEXT KAKENHI Grant-in-Aid for Transformative Research Areas A Extreme Universe No.21H05184 (N.I.), No.21H05186 (A.I. and K.M.), 
and 21H05182 (N.I. and A.I.). 

\appendix
\section{The field equations for $f$ and $g$}
\label{Appendix:A}
\begin{align}
\label{eq_f}
& 3u^2fgh(uh'-3h)f'+3h[r_+^2h(\omega+2k\Omega)^2-2ufg\{2k^2-(\epsilon_k^2+\epsilon_{k+1}^2)(2h-uh')\} ]\phi^2
\nonumber \\
&+12u^3f^2gh^2\phi'^2-r_+^2fgh(2uh'-3h)V(|\phi|)-3r_+^2u^2fh^3\Omega'^2 \nonumber \\
&+3ufg\{h^2(-8+3f+5h)+u(4-3f-2h)hh'+u^2fh'^2 \}=0, 
\end{align}
\begin{align}
\label{eq_g}
& 3u^2f^2h(uh'-3h)g'-6h[r_+^2h(\omega+2k\Omega)^2+ufg\{-4k^2
+(\epsilon_k^2+\epsilon_{k+1}^2)(h-uh') \} ]\phi^2 \nonumber \\
&-24u^3f^2gh^2\phi'^2+2r_+^2ufghh'V(|\phi|)+6r_+^2u^2fh^3\Omega'^2 \nonumber \\
&-3ufg\{4h^2(h-1)-uh(-4+f+2h)h'+u^2fh'^2 \}=0. 
\end{align}

\section{Expansions of the field equations}
\label{Appendix:B}
Under the conditions, (\ref{corotating_cond}) and (\ref{cond_Omega_phi_0}), the field Eqs.~(\ref{eq_f}) and (\ref{eq_g}) are
expanded around the horizon $u=1$ as
\begin{align}
\label{expansion_fg_phi0}
& 3u^2(uh'-3h)f'=
3\{h_0(8-5h_0-2h_1)-2(2h_0+h_1)(\epsilon_k^2+\epsilon_{k+1}^2)\phi_0^2 \nonumber \\
&+4(h_1+k^2\phi_0^2)\}-(3h_0+2h_1)r_+^2V(|\phi_0|)+O(1-u) \nonumber \\
&=F_0+O(1-u), \nonumber \\
& \frac{3u^2f(uh'-3h)g'}{g}=6\{h_0(-2+2h_0+h_1)+(h_0+h_1)(\epsilon_k^2+\epsilon_{k+1}^2)\phi_0^2-2(h_1+2k^2\phi_0^2)\}
\nonumber \\
&+2h_1r_+^2V(|\phi_0|)+O((1-u)) \nonumber \\
&=G_0+O(1-u).  
\end{align} 
The r.~h.~s.~of Eqs~(\ref{expansion_fg_phi0}) should be zero if the metric is $C^2$. This yields the solution~(\ref{sol_phi0}).  

\section{The first law of the extremal Myers-Perry AdS black hole}
\label{Appendix:C}
In the extremal Myers-Perry AdS black hole, the energy~(\ref{Q_total}) and the angular momentum 
$J$~(\ref{angular_momentum}) are evaluated from the metric~(\ref{Myers-Perry}) 
under the condition $\alpha=0$ as 
\begin{align}
\label{MP_M_J}
& Q_{tot}=\pi^2\ell^2\hat{r}_+^2(8\hat{r}_+^4+13\hat{r}_+^2+6), \nonumber \\
& J=8\pi^2\ell^3\sqrt{1+\frac{1}{2\hat{r}_+^2}}\left(1+\frac{1}{\hat{r}_+^2}\right)\hat{r}_+^6. 
\end{align}
By taking the deviation $\hat{r}_+\to \hat{r}_++\delta$, we obtain 
\begin{align}
\delta Q_{tot}=\frac{\sqrt{1+{1}/{2\hat{r}_+^2}}}{\ell}\delta J=\Omega_h\,\delta J, 
\end{align}
where $\Omega_h=\Omega(1)$ is the angular velocity of the rotating Myers-Perry black hole. 
Since the temperature of the extremal Myers-Perry AdS black hole is zero, this is the first law of the 
extremal Myers-Perry AdS black hole.


\end{document}